\documentclass[aps,prb,twocolumn,showpacs,amsmath,amssymb,footinbib]{revtex4-1}
\usepackage{hyperref}
\usepackage{graphicx}
\usepackage{bbm}
\usepackage{color}

\begin{document}
\title[]{Effective crystal field and Fermi surface topology: a comparison of 
 $d$- and $dp$-orbital models}
\author{N. Parragh$^1$, G. Sangiovanni$^1$, P. Hansmann$^2$,
  S. Hummel$^3$, K. Held$^3$, and A. Toschi$^3$}

\affiliation{
${}^1$Institut f\"ur Theoretische Physik und Astrophysik, Universit\"at W\"urzburg, Am Hubland, D-97074 W\"urzburg, Germany \\
${}^2$ Centre de Physique Th\'eorique, Ecole Polytechnique, CNRS-UMR7644, 91128 Palaiseau, Paris, France \\ 
${}^3$ Institut f\"ur Festk\"orperphysik, Technische Universit\"at Wien, 1040 Vienna, Austria}


\begin{abstract}
The effective crystal field in multi-orbital correlated materials can be either enhanced or reduced by electronic correlations with crucial consequences for the topology of the Fermi surface and, hence, on the physical properties of these systems.
In this respect, recent 
local density approximation (LDA) plus dynamical mean-field theory (DMFT)
studies of Ni-based heterostructure
have shown contradicting results,
 depending on whether  the less correlated $p$-orbitals are included or not.
We investigate the origin of
this problem and identify the key parameters controlling
the Fermi surface properties of these systems.
Without the $p$-orbitals the model is quarter filled, while the $d$ manifold moves rapidly towards half-filling when the $p$-orbitals are included. This implies that the local Hund's exchange, while rather unimportant for the former case, can play a predominant role in controlling the orbital polarization for the extended basis-set by favoring the formation of a larger local magnetic moment.
\end{abstract}

\maketitle

\section{Introduction}

Correlated electronic systems display some of the most fascinating phenomena in solid state physics. 
One of their typical characteristics is a strong sensitivity to
 small changes of external control parameters. Hence, a precise understanding
of the underlying physics and of the pivotal
parameters controlling the observed phenomenology represents a crucial
goal in contemporary condensed matter research, also in light of 
possible applications beyond the purely academic context.

The intrinsic complexity of  many-body physics
prevents an exact ab-initio theoretical description 
of correlated materials. In fact, even one of the most basic models for
electronic correlations, i.e. the Hubbard model\cite{Hubbard} 
where only the local part of the Coulomb
interaction is retained, cannot be exactly solved in the relevant cases of two or three dimensions.
However, a great step forward in the theoretical analysis of
electronic correlation in condensed matter  was achieved in the last
 decades by means of  dynamical mean-field theory (DMFT)\cite{Metzner1989,DMFTrev},
and, for realistic material calculations, by its merger with ab-initio
density functional approaches (LDA+DMFT)\cite{LDADMFT}.

From the theoretical point of view, DMFT-based methods can be viewed 
as a quantum extension of the classical mean-field approaches. 
Hence, the application of DMFT implies neglecting all non-local spatial correlations albeit allowing 
for a very accurate (and non-perturbative) treatment
of the most relevant part of the electronic correlations stemming
from the  (single or multi-orbital) Hubbard interaction
i.e., their purely local part. The ability of DMFT to capture these
local quantum fluctuations is one of the keys
behind its success in addressing many open questions in the physics of
strongly correlated materials. Among those, we recall the pioneering
DMFT description of the Mott\cite{MH} metal-insulator transition (MIT) in
V$_2$O$_3$\cite{DMFTrev,Held2001}, of the $\delta$ phase of Pu\cite{Savrasov2001},
of the correlation effects\cite{Licht2001} in Fe and Ni, of the volume collapse in Ce\cite{Held2001_B},  
of the unconventional pairing mechanism of superconductivity in
fullerenes\cite{Capone2002}; most recently DMFT has been also successfully applied
to the analysis of the occurrence of kinks in the
self-energy\cite{Byczuk2007}  and in the specific heat\cite{PRLkinks}  of particular
vanadates, such as  SrVO$_3$, and LiV$_2$O$_4$, as well as of the spectral
and magnetic properties\cite{Pnictides_donly,Pnictides_donly2} of Fe-based
superconductors.

Furthermore, it should also be recalled here, that
DMFT is a very flexible scheme, whose application is possible also beyond the
standard case of bulk correlated systems. In fact,  DMFT-based methods have been
recently used to study  correlated nanoscopic\cite{nanoDMFT} and
hetero-structures\cite{Potthoff1999,Okamoto2008,Hansmann2009}.
For the latter case, we want to focus here,
in particular, on the theoretical predictions for the Fermi-surface
properties of layered Ni-based heterostructures.  The application of
LDA+DMFT to this problem, and more specifically, to the case of a $1\!:\!1$
layered LaNiO$_3$/LaAlO$_3$ heterostructure has raised a considerable
interest, as the DMFT results of Refs. \onlinecite{Hansmann2009,Hansmann2010} clearly prospect the
possibility to drive the Fermi surface ``topology'' of these materials
very close to the one of the high-temperature superconducting
cuprates. In fact, the electronic structure  in the bulk Nickelates, such as
R$_{1-x}$Sr$_x$NiO$_4$,  is typically characterized by {\sl two} bands crossing
the Fermi level\cite{Uchida2011}, which arise from the two $e_g$-orbitals of Ni
(the ``planar'' $x^2 -y^2$ and ``axial'' $3z^2-r^2$-orbital). However, by growing heterostructures
with planes of LaNiO$_3$ intercalated with insulating planes of
LaAlO$_3$ and on substrates providing an epitaxial strain, such as SrTiO$_3$ or PrScO$_3$,
the energetic configuration of the $3z^2-r^2$ will be correspondingly disfavored\cite{Khal2008},
which corresponds to a (by our definition positive) crystal field splitting $\Delta_{CF}^d = \varepsilon_{3z^2-r^2} - \varepsilon_{x^2-y^2} > 0$ among the two
Ni $e_{g}$-orbitals. In this situation,  LDA+DMFT calculations have shown that
the inclusion of the correlation effects will always increase the
original (LDA) crystal field splitting between the
two $e_g$-orbitals\cite{noteV2O3}, leading, eventually, to a significant change of the ``topology''
of the Fermi surface, i.e. to a situation in which only
{\sl one} band, with predominant $x^2\!-\!y^2$-character crosses the Fermi level.
Hence, according to these LDA+DMFT calculations, in Ni-based heterostructures it would be possible
to ``artificially'' realize the electronic configuration of the high-temperature
superconducting cuprates (i.e., single, almost half-filled orbital with
$x^2\!-\!y^2$ symmetry close the Fermi level), e.g.  by modulating the
strain through changing the substrate or the insulating layer.
 As the control of the low-Fermi surface
properties represents an essential ingredient for  novel, alternative, 
realizations of  high-temperature superconductivity, the importance of
having  highly accurate LDA+DMFT predictions becomes a crucial
factor for engineering new materials. 

In contrast to these impressive applications of DMFT-based
methods, one important aspect should be stressed here: 
The LDA+DMFT procedure requires as an important step a downfolding to a chosen energy window around the Fermi energy $\varepsilon_F$. For most of the above mentioned studies, this window included only a few bands around $\varepsilon_F$ of dominant 3d character. In such basis for the transition metal oxides the hybridization of 3d states and oxygen ligand 2p states is included \emph{implicitely} in the effective bands. 
In the recent past, however, it has become customary to include the oxygen 2p states \emph{explicitely}, i.e. to downfold to a larger energy window.
In most of the cases, these additional ($p$-) orbitals were
way more extended than the correlated ones (e.g. $d$), and, hence, the local
Coulomb interaction between electrons occupying these additional orbitals
($U_{pp}$) and between electrons on different manifolds
($U_{pd}$) was either completely neglected, or, in some exceptional cases, treated at the Hartree level\cite{Thesis_PH}.  Despite this approximation, it is quite intuitive to expect that the validity of  LDA+DMFT calculations performed
in an enlarged (say: $dp$) basis-set is more general than the
corresponding one in the restricted $d$ manifold. In fact, quite
generally:  (i)  performing a renormalization (Wannier\cite{Wannier} projection, NMTO\cite{NMTO} downfolding, etc.) on 
an enlarged basis-set allows for a better localization of the
orbitals of the correlated manifold (as the $dp$-hopping processes are
now explicitly included in the model); and (ii) the possibility of
describing explicitly charge-transfer processes between the $d$ and
$p$-orbitals makes the theoretical modeling evidently closer to the
actual material physics\cite{noteMaurits}.

Notwithstanding these quite general arguments, the improvement of 
LDA+DMFT calculations on enlarged $dp$ basis-sets 
w.r.t. the ones restricted  to effective $d$-only basis, is not always apparent. 
In fact, there are cases for which a treatment on a larger basis set renders the comparison with experiment to be worse!
Without attempting to give
a complete review here, we recall that LDA+DMFT calculations including $p$-orbitals have improved the descriptions of the insulating behavior of
NiO\cite{Kunes2007} and of the MIT in
NiS$_2$\cite{Kunes2009}  w.r.t. $d$-only calculations\cite{PRB_pirite}. 
Also quite accurate results have been obtained for one-
and two-particle properties of cobaltates (such as SrCoO$_3$\cite{PRLwithJan}
and LaCoO$_3$ \cite{PRB_LaCoO3}) and in several studies\cite{Pnictides_dp}
of the well-known class of iron-pnictides and calchogenides. 

In contrast to the aforementioned successful applications 
of $dp$ calculations, in
other, equally important, cases the $dp$ LDA+DMFT results
are in
partial or total contradiction with the $d$-only calculations, 
and/or with the experimental findings: 
No MIT  in 
V$_2$O$_3$ was found up to unrealistically large values
of the Coulomb interaction
if the Oxygen $p$-orbitals are included
in LDA+DMFT calculations \cite{Karstenslides}.
More recently also
the Mott-Hubbard insulating phase of La$_2$CuO$_4$
 and  LaNiO$_3$ was reported to
be missing
in the $dp$ framework\cite{Lucamillis},
while these materials are found to be insulating 
in $d$-only calculations for plausible values of the $dd$ interaction.
 These discrepancies between
$d$-only and $dp$ calculations regarding
 the Mott-Hubbard MIT have already raised
a discussion in the recent literature. 
In Ref. \onlinecite{Lucamillis}, non-local correlations to be included beyond 
DMFT\cite{beyondDMFT}  have been considered  as a cause of the discrepancy.  
In fact, a major role of spatial correlations in determining the onset of insulating phases is quite likely, especially in the case of two-dimensional cuprates.

However, there are also other discrepancies, whose discussion will be
the object of the present work, which can be hardly attributed to the
effects of non-local correlations. These discrepancies are observed for
systems of more than one correlated $d$-orbital and in broad
parameter regimes (including high temperatures), where effects beyond
DMFT should not play a crucial role. In particular, a striking
disagreement between $d$-only and $dp$ calculations was reported for
the above-mentioned case of Ni-based heterostructures. In fact,
LDA+DMFT calculations performed including also the Oxygen $p$-orbitals have shown\cite{Hanmillis} exactly the opposite trend w.r.t. the previous ones: 
Even in presence of a favorable crystal field splitting $\Delta_{CF}^d=
\varepsilon_{3z^2-r^2} - \varepsilon_{x^2-y^2} > 0$  
at the starting (LDA) level,  the net effect of the Hubbard interaction was {\sl  always}
 to reduce the orbital polarization by  filling back  the $3z^2-r^2$-orbital,  which would prohibit  \emph{de facto} any possibility of realizing the
 cuprate conditions  for the onset of an unconventional
 superconductivity.

This second kind of discrepancies between LDA+DMFT performed with
different ($d$-only vs. $dp$) basis-sets  well illustrated by the
contradicting results for the Ni-based heterostructures, raises a
quite general question about the proper use and interpretation of
the growing number of LDA+DMFT calculations on extended basis
sets. The correct determination of the orbital polarization
and Fermi surface properties is of great importance
for  future calculations of increasingly complex materials.
 
In this paper  we aim at understanding
the relations between the results of existing LDA+DMFT calculations on different basis-sets for the
Fermi surface properties of multi-orbital systems and, ultimately,
the origin of the qualitative discrepancies observed by following the
standard implementation of the algorithms in different basis-sets.
For this purpose,  it is of primary importance to disentangle the main, qualitative,
trends from the specific features of a selected
case. 
Hence, we will study model Hamiltonians for different crsytal field splittings starting from a dp-basis (4 bands). Subsequently, we perform a downfolding to an effective d-only basis and compare DMFT ressults of both cases.
This procedure will capture the above-mentioned
discrepancy in the prediction of Fermi surfaces of correlated
multi-orbital systems and allow for a systematic study of
this problem in the context of Ni-based materials.


The scheme of the paper is the following: In Sec. II we introduce
the models. In Sec. III, we analyze the 
disagreement in the calculation of the orbital occupations and the shape of the Fermi surfaces  between
 $d$-only  and $dp$ model 
 for a fixed set of interaction parameters. At the end
of Sec. III, we also provide an  analysis of the  role of Hund's exchange $J$. In Sec. IV 
we study the origin of the observed inconsistencies by analyzing 
the dependence on the $d$-electron-density, 
and we discuss its possible relation with the crossover\cite{GeorgesPROC} from
high-spin (Hund's regime) to low-spin (CF regime). 
Finally, in Sec. V we summarize our results.

\vskip 5mm

\begin{figure*}[t!]
         \centering\includegraphics[width=0.65\textwidth,angle=0]{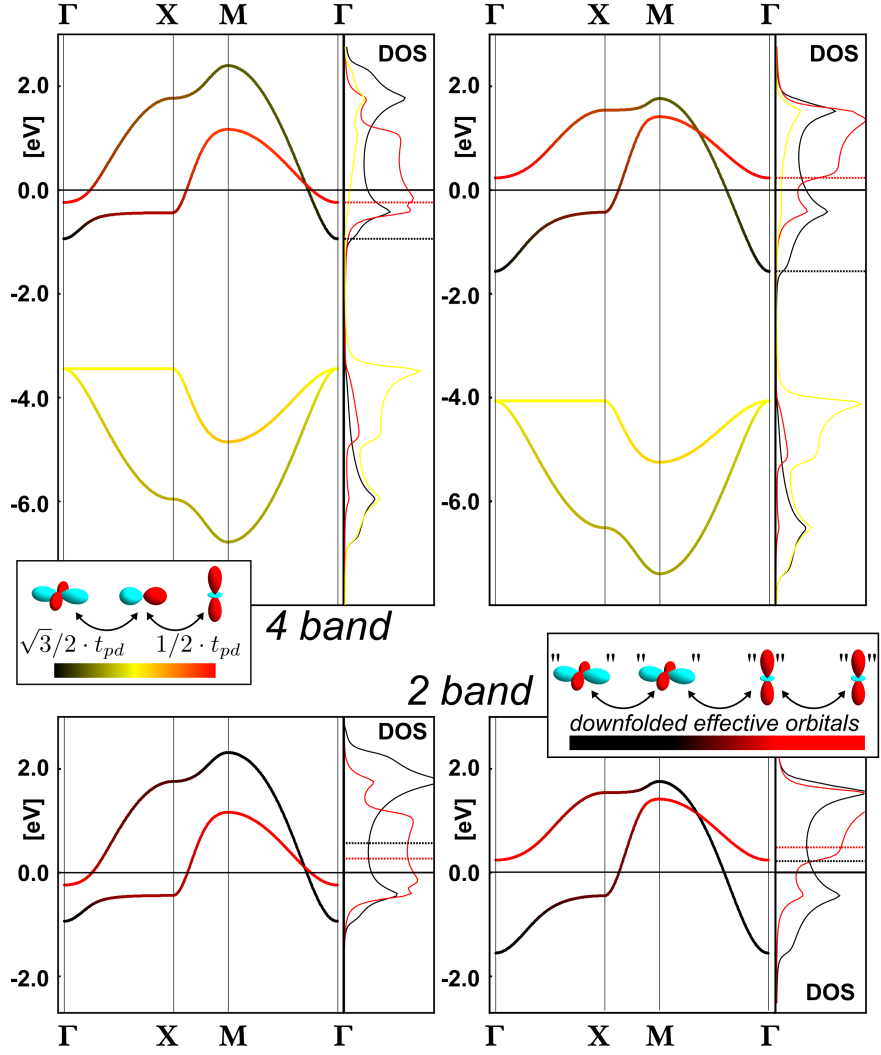}
        \caption{\label{fig01} Band-structure and DOS of the
          four-band $dp$ model (upper panel) and the corresponding downfolded two-band $dp$ model (lower panel) for two different values of the crystal field splitting $\Delta_{CF}$ between the $e_g$-orbitals of the $dp$ Hamiltonian, Eq. \ref{eq:dpmodel} . Specifically these are 
$\Delta_{CF}= 0.7$eV (left panels) and $1.7$eV (right panels), corresponding to
a local energy splitting $\Delta_{CF}^d = -0.53$eV and $0.37$eV respectively in the $d$-only model. The orbital character is denoted by the following
color-coding: black for the first $d$-orbital
          ($x^2\!-\!y^2$), red for the second  ($3z^2-r^2$),  and yellow for the $p$-orbitals. In the inset, the hopping processes for the two models are sketched.}
\end{figure*}

\section{Models and Methods} 

In this section, we illustrate the simplified model Hamiltonians chosen to analyze the discrepancies between the existing LDA+DMFT calculations on different  ($d$-only vs. $dp$) basis-sets for the prediction of 
the orbital polarization and the Fermi surface properties of correlated multi-orbital systems. Let us stress that, in this work, we aim at a basic understanding of the general physics underlying the contradicting outcomes of realistic calculations, rather than discussing the case of a specific material.
As a guidance for the choice of the model, we recall (see Sec. I) that important 
inconsistencies in the theoretical predictions emerged already when
considering the presumably simple case of two correlated (namely, $e_g$-)\cite{note1} orbitals only, such as in the case of bulk Nickelates\cite{Uchida2011} and Ni-based heterostructures\cite{Hansmann2009,Hansmann2010,Hanmillis}.

Hence, we will introduce here two models (distinguished for the different basis-sets), built with tetragonal symmetry, allowing for the simplest realization of the above-mentioned situation of two correlated orbitals (identified with the two $e_g$ $d$-orbitals) in the presence of a mimal number of extended ligand orbitals (henceforth associated to $p$-orbitals). The models are inspired, to some extent, by the realistic calculations of nickelate heterostructures, i.e. with a focus on the orbital polarization in a
quasi two-dimensional geometry. However, we will not consider here spatial correlations beyond DMFT\cite{beyondDMFT}, the ``dimensionality''  entering explicitly only via the $k$-summations defining the local quantities. Therefore, the results will depend mainly on the ``effective'' crystal field splitting, on the orbital occupation and on the electronic kinetic energies. Furthermore, the general trends emerging from our DMFT calculations are found to be stable w.r.t. the specific parameter choice of the $dp$ and $d$-only models we are going to consider.  Hence, our study will provide useful indications to identify and clarify the global trends of the multi-orbital physics of this class of systems, {\sl at the DMFT level}.   

Specifically, the model Hamiltonians considered will be the following:
 
\begin{enumerate}
 
\item The one-particle model Hamiltonian for the larger basis-set, from which we start conceptually, is a $dp$ four-band model (Fig.\ref{fig01} upper panels), that includes explicitly two correlated $d$-orbitals (e.g., the Nickel $3d$-$e_g$ states), and two ligand (e.g., Oxygen $2p$ states).

\item The model for the smaller basis-set is the (L\"owdin) downfolded\cite{Loewdin, Juelichschool} version of the first one on the two correlated (e.g. $3d$-$e_g$) orbitals:  It corresponds to the two-orbital $d$-only band-structure, shown with the corresponding orbital character in  Fig.\ref{fig01} (lower panels). 

\end{enumerate}

Note that, due to the downfolding, the band structure of the two models at the Fermi level are identical by construction. Yet the associated orbital character is different.

\subsubsection*{dp four-band model} 

While DMFT calculations are often performed first in the smallest possible basis-set, for the definition of our one-particle model Hamiltonians it is conceptually more logical to start from a larger ($dp$) basis-set. In our case, this includes two correlated $d$-orbitals and two ligand $p$-orbitals. The specific dispersion considered corresponds to a quasi two-dimensional geometry of the model, which can be thought of as one $p_x$- and one $p_y$-orbital on each ligand site. The overlap between the ligand orbitals and the two $e_g$ orbital gives the largest contribution to the hopping processes ($t_{pd}$). Hence, for the sake of simplicity, the smaller direct hopping among $d$-orbitals is neglected ($t_{dd}=0$). Furthermore, in our one-particle Hamiltonian the $d$ and the $p$ manifold are kept well distinguishble, in that a large enough $dp$ splitting is assumed (see below for details).
These choices correspond to the minimal $dp$ model with a realistic $dp$ configuration in a (quasi two dimensional) cubic/tetragonal symmetry.

The specific values of the parameters, e.g., the $dp$-hopping amplitude ($t_{pd}$), on-site $d$/$p$ energies, and crystal field splitting among the $d$-orbitals ($\Delta_{CF}$) have therefore been chosen to obtain a reasonable bandwidth for the bands close to the Fermi level. Our choice also ensures that the positive/negative values assumed by the splitting $\Delta_{CF}^d$ between the two downfolded $d$-orbitals in the small basis-set are plausible. This way we build up a one-particle low-energy Hamiltonian similar to the 4-band cuprate Hamiltonian derived in Ref. \onlinecite{OKAmodels}, whereas the role of our ``axial'' orbital $3z^2-r^2$ is played by the $4s$-orbital of Cu. Formally, our $dp$ Hamiltonian in momentum space reads:
\begin{widetext}
 \begin{equation}
\hspace{-2cm}
 H^{\rm 4b}_\mathbf{k}=
\left(
\begin{array}{cccc}
 0 & 0 & i \sqrt{3} t_{pd} \sin\Big(\frac{k_x}{2}\Big) & -i \sqrt{3} t_{pd} \sin\Big(\frac{k_y}{2}\Big) \\
 0 & \Delta_{CF}  & i t_{pd} \sin\Big(\frac{k_x}{2}\Big) & i t_{pd} \sin\Big(\frac{k_y}{2}\Big) \\
 -i \sqrt{3} t_{pd} \sin\Big(\frac{k_x}{2}\Big) & -i t_{pd} \sin\Big(\frac{k_x}{2}\Big) & \epsilon_p & 0 \\
 i \sqrt{3} t_{pd} \sin\Big(\frac{k_y}{2}\Big) & -i t_{pd} \sin\Big(\frac{k_y}{2}\Big) & 0 & \epsilon_p
\end{array}
\right),\label{eq:dpmodel}
\end{equation}
\end{widetext}
where the lattice spacing $a=1$, $t_{dp}=1.8$eV and $\Delta_{CF}= [0.7, 1.7]$eV ($\Delta_{CF}^d= [-0.53, +0.37]$eV) for the reason explained above. The on-site energy of the $p$-states was fixed to $\epsilon_p=-2.5$eV similar to the  position of the $p$-bands in the nickelate systems. As for the Fermi level, the model is assumed to have typically overall filling of $5$ electrons. This would correspond, ideally, to the situation of filled ligand bands and to a quarter-filled correlated two-band manifold at the Fermi level.

In the lower panels of Fig.~\ref{fig01}, we show the band-structure of the four-band $dp$ model corresponding to the two extreme values of $\Delta_{CF}$ we considered. In the band-structure and DOS plots the black/red color encodes the $d$-character ($x^2\!-\!y^2$/$3z^2-r^2$, respectively), and the yellow color encodes the $p$-character. 

\subsubsection*{d-only two-band model} 

The $d$-only model has been obtained by means of a L\"owdin downfolding of the $dp$ Hamiltonian, Eq.\ \ref{eq:dpmodel} which means that their respective electronic structure at the Fermi level is identical by construction.
The overall result of the downfolding procedure is reported in the corresponding orbitally-resolved bandstructure for the two-band model in Fig.\ \ref{fig01} (bottom panels). Specifically, the bandstrucuture and DOS of Fig. \ref{fig01} have been obtained by downfolding the $dp$ Hamiltonians with the lowest and the highest value of $\Delta_{CF}$ (and $\Delta_{CF}^d$), respectively. The orbital occupation of the $d$-only model has been fixed to $n_d=1$ (quarter-filling).

In Fig. \ref{fig01} the color code denotes, similarly as before, the orbital
character: black and red represent the downfolded $x^2\!-\!y^2$- and
$3z^2-r^2$-orbital, respectively. At a closer inspection of the figure, we note
that, while locally the two $e_g$-orbitals are eigenstates of the tetragonal
point group and do not hybridize (pure black/red color of the DOS), they do
obviously hybridize non-locally along certain directions (dark-red color, e.g.
at the X point of the Brillouin zone). Furthermore, we observe that the overall
width (as well as the associated kinetic energy) of the $3z^2-r^2$ DOS is
smaller than the $x^2\!-\!y^2$ one, a typical feature of anisotropic materials
and/or heterostructures with a quasi two-dimensional  hopping geometry with
suppressed hopping along the $c$-axis. More importantly for our purposes, we
stress how the different original values of $\Delta_{CF}$ in the $dp$
basis-sets correspond to situations with negative/positive difference between
the center of mass of the non-interacting DOS of our $d$-only model.
Specifically, this means that the important parameter $\Delta_{CF}^d =
\varepsilon_{3z^2-r^2} - \varepsilon_{x^2-y^2}$, accounting for the on-site
energy difference\cite{noteDeltaCFd} between the two $e_g$ states {\sl in the
downfolded model}, will be varying in the range of $[- 0.53, 0.37]$ depending on the chosen value of $\Delta_{CF}$ in the starting $dp$ Hamiltonian.  
We stress here, that $\Delta_{CF}^d$ can be interpreted, to a certain extent, as ``crystal field'' within the $d$-only manyfold. However, by adopting such a ``rough'' definition, one should keep in mind that $\Delta_{CF}^d$ is, in fact, a ``ligand field'' splitting, originated by an electrostatic Madelung potential {\sl and} a $dp$-hybridization splitting.

Before discussing some technical aspects of our DMFT implementation in the next subsection, we recall that related two-band model Hamiltonians, but {\sl without} hybridization between the $d$-orbitals, have been studied with DMFT in Ref. \onlinecite{GeorgesPROC}, with reference to the cases of BaVS$_3$ and Na$_x$CoO$_2$. \\

\subsubsection*{DMFT algorithm in extended basis-sets}

The DMFT algorithm applied to a chosen correlated orbital-subspace, possibly derived from ab-initio calculations, has already become a quite standardized procedure (for details we refer the reader to \onlinecite{LDADMFT}). However, in consideration of the discrepancies appearing when extending the basis-set by including less-correlated ligand $p$-orbitals in the self-consistent DMFT loop, we will explicitly recall here some technical aspects of the DMFT algorithmic procedure usually adopted in the cases of the extended $dp$-basis-sets.

Extended models including ligand $p$-states explicitly have a
structure like our 4-band Hamiltonian (\ref{eq:dpmodel}).
 Locally, i.e. integrated over the Brillouin zone, such Hamiltonian has the form:

\begin{equation}\label{dpHAM}
 H^{\rm loc.}_{\rm full}(R=0)=
 \left(
 \begin{array}{cc}
   H^{\rm loc.}_{dd} & H^{\rm hyb.}_{dp}\\ 
   \left(H^{\rm hyb.}_{dp}\right)^\dagger & H^{\rm loc.}_{pp} \\
 \end{array}
 \right)
\end{equation}
The local basis is typically chosen in a way that the $H_{dd}$ and
$H_{pp}$ blocks can be made internally diagonal after {\bf k}--integration so that the
states can be labeled by a good \emph{local} quantum number in the
respective subspaces, such as the crystal field labels (see DOS plots
in Fig.~\ref{fig01}).   In such a basis, the local Coulomb ($U$)
matrix of the interacting part of
the Hamiltonian is then defined (in its SU(2)-invariant ``Kanamori'')
form as 
\begin{widetext}
\begin{equation} 
\begin{array}{rl}
H_{loc}=&\;\sum_{a} U n_{a,\uparrow} n_{a,\downarrow}\nonumber\\
&+\sum_{a>b,\sigma} \Big[U' n_{a,\sigma} n_{b,-\sigma} +  (U'-J) n_{a,\sigma}n_{b,\sigma}\Big]\nonumber\\
&-\sum_{a\ne b}J(d^\dagger_{a,\downarrow}d^\dagger_{b,\uparrow}d^{\phantom{\dagger}}_{b,\downarrow}d^{\phantom{\dagger}}_{a,\uparrow}
+ d^\dagger_{b,\uparrow}d^\dagger_{b,\downarrow}d^{\phantom{\dagger}}_{a,\uparrow}d^{\phantom{\dagger}}_{a,\downarrow} + h.c.).
\end{array}
\label{Kanamori}
\end{equation}
\end{widetext}
for the $d$-orbital sector. Here, $U$ denotes the interaction parameter between two electrons in the same
$d$-orbital, $U'$ the interaction between electrons on different
$d$-orbitals and $J$ is the Hund's coupling; $a,b$ index the two orbitals and $\sigma$ the
spin. Note that the specific interaction values for the DMFT
calculations (see next section for details) have been chosen in order
to reproduce a typical correlated metallic 
situation. Due to the stronger localization of the
$d$-orbitals in the $dp$ models, the corresponding values of the local
interaction on the $d$-orbitals have been correspondingly enhanced (we assumed
here a factor two for the parameter $U'$).  We recall that -- for the
$dp$ case --  the multi-orbital 
Hubbard interaction could include, besides on-site $d$ and on-site
$p$ interactions, also possible $dp$ interactions.
This is, however, not the topic of the present study. 
Furthermore, for the $dp$ model in LDA+DMFT we have to face the so-called
problem of \emph{double-counting correction} (DC)\cite{LDADMFT,DC_ref,DC_pot}:
Unlike for a $d$-only model, this does not correspond to a simple total energy shift
and, hence, cannot be ``absorbed'' in the chemical potential. The DC for the
$dp$ models corresponds to a renormalization of the energy difference between $d$
and $p$ states. For our models we have used the DC suggested by Anisimov
\cite{DC_ref}, given by:
\begin{equation}
\label{AnisimovDC}
\begin{array}{rcl}
      \bar{U}_{dd}& =& \big[ U + U'  (N_d-1) + (U'-J)  (N_d-1) \big] / \left(2  N_d -1 \right)\\
 \\
      \Delta_{DC}& =& \bar{U}_{dd} (\sum_d n_d^{LDA} - \frac{1}{2}),
\end{array}
\end{equation}
where $N_d$ is the number of $d$-orbitals, and $\bar{U}_{dd}$ is the 
average local interaction between these $d$-orbitals; 
$n^{LDA}_d$ denotes the occupation of the $d$-orbitals as calculated
from the {\bf k}-resolved non-interacting (LDA) Hamiltonian discussed above.

The self-consistent DMFT loop, which includes the solution of an
Anderson impurity problem for the correlated subspace at each step is
done as follows:

\begin{enumerate}
\item  The first step is the calculation of the $\mathbf{k}$--integrated Green function on the
  \emph{full} $dp$ basis-set:
\begin{eqnarray}\label{C4kint}
     G^{{\rm loc. }}_{\rm full}(\omega)  = && \frac{1}{V_{{\rm BZ}}}  \int_{{\rm BZ}}d^3k  \nonumber\\
&& \big[(\omega+\mu)\mathbbm{1}- H^{\rm 4b}_\mathbf{k}-\Sigma_{\rm full}(\omega)\big]^{-1}
  \end{eqnarray}
where $G$, $\Sigma_{\rm full}$, and $H_\mathbf{k}$ are matrices in the $dp$-basis.\\
\item Next, we extract the $dd$-block of the local Green function:\\ 
\begin{equation}\label{C4dyson}
G^{{\rm loc. }}_{dd}(\omega)=\left\{ G^{{\rm loc. }}_{\rm full}(\omega) \right\}|_{dd-{\rm block}}, 
\end{equation}

\noindent
i.e. we project it onto the $d$-subspace. We stress
here, that due to the $dp$ hybridization encoded in
the Hamiltonian $H^{\rm 4b}_\mathbf{k}$ and the inversion of
Eq.~\ref{C4kint} the information about the $p$-ligands is not lost but
captured by $G^{{\rm loc. }}_{dd}(\omega)$.\\
\item Now, in complete analogy to DMFT for $d$-states only, we calculate the
  Weiss field for the impurity model (only on the $d$-subspace):\\
\begin{equation}\label{C4WeissField}
\left[{\cal G}^{0}(\omega)\right]^{-1}=\left[G_{dd}(\omega)\right]^{-1}+\Sigma_{dd}^{\rm DMFT}(\omega), 
\end{equation} 

\noindent
where $\Sigma_{dd}^{\rm DMFT}(\omega)$ denotes the DMFT $dd$ self-energy.\\ 
\item With ${\cal G}^{0}(\omega)$ we solve the auxiliary impurity problem (see below), obtain a new impurity Green function and a new self-energy for the $d$-orbitals. \\  
 Finally the self-consistent loop is closed by comparing
both the new and the old $\Sigma_{dd}^{\rm DMFT}(\omega)$ \emph{and} the new and
the old $d$/$dp$-density and iterating until convergence. 
\end{enumerate}

As impurity solver for our DMFT calculations, we have used the
continuous time quantum Monte-Carlo (QMC) algorithm in the 
hybridization expansion (CT-HYB)\cite{Werner2006,Gull2011}, which
allows also  for the treatment of the spin-flip and pair-hopping terms
of our Kanamori Hamiltonian (\ref{Kanamori}). In the CT-HYB the
time evolution is calculated using the local interaction, which
makes the local Hilbert space  growing exponentially with the number of orbitals. 
While for density-density interactions  this problem
can be mitigated along the line of \cite{Werner2006,Haule2007},
for the more realistic Kanamori interaction a set of quantum
numbers, called PS, leads to a more efficient algorithm 
\cite{Parragh2012}. We have used this to perform all calculations presented in this paper with the full SU(2)-symmetric Hamiltonian.

\section{Results: $\mathbf{d}$ \emph{vs} $\mathbf{dp}$ calculations at ``quarter-filling''}

In this section, we compare our DMFT results obtained in the two different basis-sets. Following the chronological order of appearence of realistic LDA+DMFT calculations for these systems, we present data for the $d$-only (downfolded) basis-set first and then the corresponding ones for the original $dp$ Hamiltonian.   

We mention that our model study can be qualitatively related, depending on the initial value of the energy splitting  between the correlated orbitals ($\Delta_{CF}^d= \varepsilon_{3z^2-r^2} - \varepsilon_{x^2-y^2}$), to the physics of bulk
Nickelates\cite{Uchida2011} and of  Ni-bases heterostructures\cite{Hansmann2009,Hansmann2010,Hanmillis}.
In fact, (i) the nominal charge of these systems corresponds also to one electron in the outer two Ni-bands and (ii) the bulk Nickelates are typically characterized by negative values of $\Delta_{CF}^d$, due to the tetragonal
distortion along the $z$-axis. In the Ni-based heterostructures instead, the localization effects in the $z$ direction, as well as the epitaxial strain due to the substrate, induce positive values for $\Delta_{CF}^d$.

\subsection{DMFT results for the (downfolded) $\mathbf{d}$-only model} \label{sec:donly}

\begin{figure}[t!]
         \centering\includegraphics[width=0.5\textwidth,angle=0]{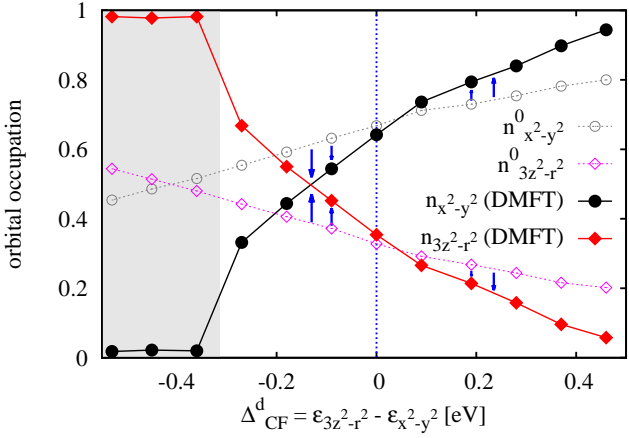} 
        \caption{\label{Fig2_dp} Orbital occupation of the $d$-only model 
       with $U'=4$eV, $J=0.75$eV ($U=5.5$eV),  $\beta=100$eV$^{-1}$ at quarter-filling ($n_d=1$) as a function of the initial crystal field
          splitting $\Delta_{CF}^d$. 
          The DMFT data (solid symbols) are
          compared with the corresponding non-interacting results (empty symbols). The arrows indicate the effect of the interaction which is essentially opposite for negative and positive $\Delta_{CF}^d$. 
The shaded region on the left indicates the onset of the Mott-Hubbard insulating phase.}
\end{figure}

\begin{figure}[t!]
         \centering\includegraphics[width=0.5\textwidth,angle=0]{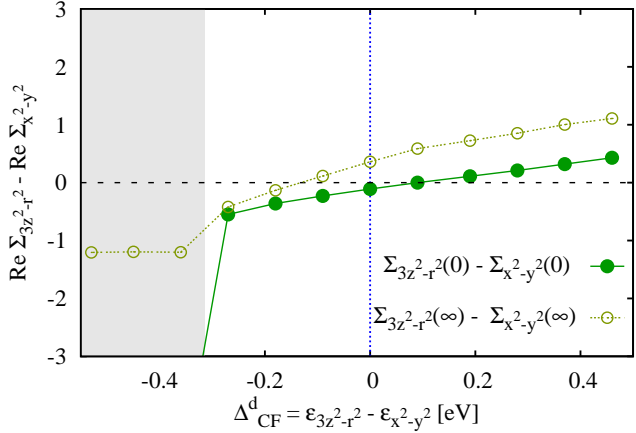}
        \caption{\label{Fig3_dp} Difference of the real part of the DMFT self-energies
          for the two orbitals of the  $d$-only model with $U'= 4$eV,
          $J=0.75$eV ($U=5.5$eV),  $\beta=100$eV$^{-1}$ at quarter-filling ($n_d=1$) extrapolated to $\omega_n \rightarrow 0$
          (solid symbols) and $\omega_n \rightarrow +\infty$  
    (empty symbol, Hartree contribution to the self-energy) as a
    function of the initial energy splitting $\Delta_{CF}^d$.
The huge enhancement of the difference between the self-energy at
$\omega_n \rightarrow 0$, and  the consequent huge energy shift of the
first orbital,  marks the onset of the Mott-Hubbard insulating phase (shaded region on the left).}
\end{figure}

In this subsection, we analyze the results for our $d$-only
two-orbital model at quarter-filling ($n_d=1$), as a function of the
initial energy splitting between the two downfolded $e_g$-orbitals  ($\Delta_{CF}^d$): the corresponding results for the orbital occupation without interaction (which would correspond to the LDA ones, in a realistic calculation) and with the
interaction (computed with DMFT) are shown in Fig. \ref{Fig2_dp}.
As mentioned in Sec. II, the interaction values have been chosen
in consideration of typical values for transition metal oxides systems: for
the $d$-only model, we adopted a value of $U'= U-2J =4$ eV, with
$J=0.75$eV ($U=5.5$eV). 

We start by briefly commenting the set of non-interacting data
shown in Fig. \ref{Fig2_dp}: They display a clear-cut
dependence on the initial value of the energy splitting of the $e_g$-orbitals: 
the orbital occupation
of the $x^2\!-\!y^2$ monotonously 
increases upon increasing values of  $\Delta_{CF}^d$,
whereas
the occupation of the $3z^2-r^2$-orbital
decreases. We
note, however, that,  since the hopping terms (and, hence, the bandwidth)
for the two orbitals are not equal, the orbital occupation curves are not
symmetric w.r.t. to $\Delta_{CF}^d$. This also implies that  the situation where the
two orbitals are equally occupied (i.e. no orbital polarization) does not occur at $\Delta_{CF}^d=0$ but, for the non-interacting
case, only for $\Delta_{CF}^d \sim -0.4$eV.

We  discuss now the effects of the Hubbard interaction on the
orbital occupations, as described by our DMFT(CT-QMC)  calculations. While the
occupation curves remain obviously asymmetric also in presence of the
interactions, from the data of Fig.  \ref{Fig2_dp} we note that, for
each orbital, the deviations w.r.t. to the non-interacting
values are strongly dependent on $\Delta_{CF}^d$ : The sign of the change in the orbital occupations appears closely connected with the sign of $\Delta_{CF}^d$. Specifically,  for $\Delta_{CF}^d > 0$ we  observe generally an enhancement (reduction) of the occupation of the first(second),
$x^2\!-\!y^2$- ($3z^2-r^2$-) orbital, while for $\Delta_{CF}^d < 0$ the trend is the opposite.

This is reflected in an analogous trend of the
change to the orbital polarization $P$, formally defined  as in Refs. \onlinecite{Hanmillis, Maurits2011}
\begin{equation}
P=\frac{n_{x^2-y^2} -n_{3z^2-r^2}}{n_{x^2-y^2} +n_{3z^2-r^2}}.
\label{Eq:pol}
 \end{equation}
as well as in the Fermi surfaces, shown in Fig. \ref{Fig6_dp}. For the latter ones, which will be discussed more extensively in the next-subsection,  our DMFT results show that, depending on the sign of the initial orbital spliting $(\Delta_{CF}^d)$, the character (as well as the typical shape) of the Fermi surface corresponding to the lower-energy orbital gets increased by the electronic interaction.     

In fact, such  trends explain the qualitatively different LDA+DMFT results previously obtained for the shape of the Fermi surface via specific $d$-only calculations for bulk nickelates\cite{Uchida2011} and Ni-based heterostructures\cite{Hansmann2009,Hansmann2010}, respectively: 
The difference in the results essentially reflects the different sign of the
initial crystal field splitting ($\Delta_{CF}^d$), as estimated by the ab-initio
calculations, whose final size ($\Delta_{eff}^d$) gets significantly magnified by the electronic interaction.

The last statement can be formalized more quantitatively through the analysis of
the  corresponding self-energies presented in Fig. \ref{Fig3_dp}: here
we show the differences between the real parts of the DMFT
self-energy of the two orbitals, i.e. $\mbox{Re} \Sigma_{3z^2-r^2}(i\omega_n)
-\mbox{Re} \Sigma_{x^2-y^2}(i\omega_n)$, evaluated in 
the limit
frequency $\omega_n\rightarrow 0$ and 
$\omega_n\rightarrow +\infty$, respectively.
 We recall that in the latter limit only 
the Hartree contributions to the electronic self-energy remain.
Hence, the difference between the self-energies
can be also explicitly written in terms of the electronic density as:
\begin{eqnarray}
 \label{Eq:Hartree} 
\hspace{-1.5cm}
\mbox{Re} \Sigma_{3z^2-r^2}(\infty) - \mbox{Re} \Sigma_{x^2-y^2}(\infty) &=& (U-
5J) \, \times \\ 
(n_{x^2-y^2}- n_{3z^2-r^2}) &=& (U -5J) \, P. \nonumber
\end{eqnarray} 
Here, the last equality only holds at quarter-filling 
where $n_{x^2-y^2} +n_{3z^2-r^2}=1$. Such a  dependence on the (final) electronic density appears
evidently in the corresponding data of Fig.\  \ref{Fig3_dp}. 
The Hartree contribution, evaluated in DMFT, changes sign \cite{note2} precisely where the
orbital occupations become equal ($P=0$), i.e. for the negative values of
$\Delta_{CF}\sim-0.15\,$eV, see Fig.\ \ref{Fig2_dp}.  By comparing this to the previous
results, it appears that the high-frequency values of the
self-energy  do not represent the ``crucial'' parameter. Instead, the trends in the orbital
polarization and, therefore, the predicted physics are controlled by the low-frequency behavior of the self-energy:
Up to the MIT the shape of the Fermi surface is determined by the ``effective'' CF splitting given by
\begin{equation}
\label{Eq:Deltaeff}
\Delta_{eff}^d = \Delta_{CF}^d + \mbox{Re} \Sigma_{3z^2-r^2}(0) - \mbox{Re} \Sigma_{x^2-y^2}(0) ,
\end{equation}
i.e. as the original crystal field 
 corrected by
the difference of the  real self-energies for
$\omega_n \rightarrow 0$.  This makes the interpretation of the
second set of data shown in Fig. \ref{Fig3_dp} very transparent: the sign of $
\mbox{Re} \Sigma_{3z^2-r^2}(0) - \mbox{Re} \Sigma_{x^2-y^2}(0) $  follows that of the original crystal field $\Delta_{CF}^d$, and confirms, from
the microscopic point of view, the picture of 
interaction effects {\sl always} magnifying the size of the original
crystal field. Of course such enhancement will
depend quantitatively on many factors: For instance, it will
be bigger when the system is more
correlated. A dramatic enhancement, in particular, is found when the Mott metal-insulator 
transition is approached, i.e. when $\Delta_{CF}^d$ approaches the
shaded area in Fig. \ref{Fig3_dp}.
Here the  low energy physics correspond to an empty
(broader) $x^2\!-\!y^2$-orbital and a half-filled 
(narrower) $3z^2-r^2$-orbital. 

However, as we will now discuss, these results are contradicted by corresponding 
calculations performed in larger basis-sets, which include also the most relevant $p$ degrees of freedom, as
it will be shown explicitly in the next subsection.

\subsection{DMFT results for the $\mathbf{dp}$ model} \label{sec:dp}

\begin{figure}[t!]
         \centering\includegraphics[width=0.5\textwidth,angle=0]{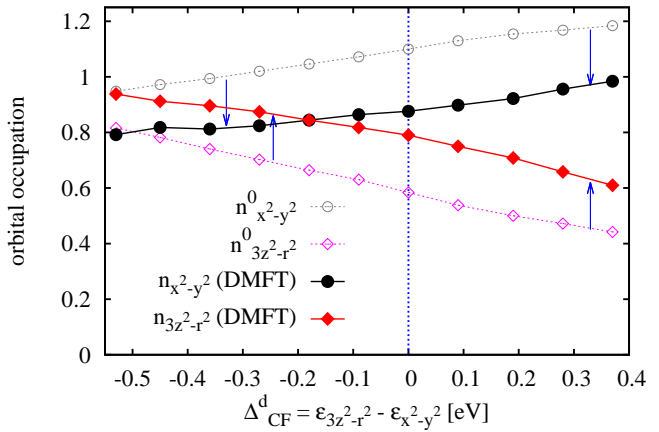}
        \caption{\label{Fig4_dp} Orbital occupation of the $d$-orbitals as a function of the initial crystal field splitting for the four-band $dp$  model with
          $U'= 8$eV, $J=1.0$eV  ($U=10$eV),  $\beta=100$eV$^{-1}$ and $n_{tot}=5$. The DMFT data (solid symbols) are compared with the corresponding non-interacting results (empty symbols).}
\end{figure}

\begin{figure}[t!]
         \centering\includegraphics[width=0.5\textwidth,angle=0]{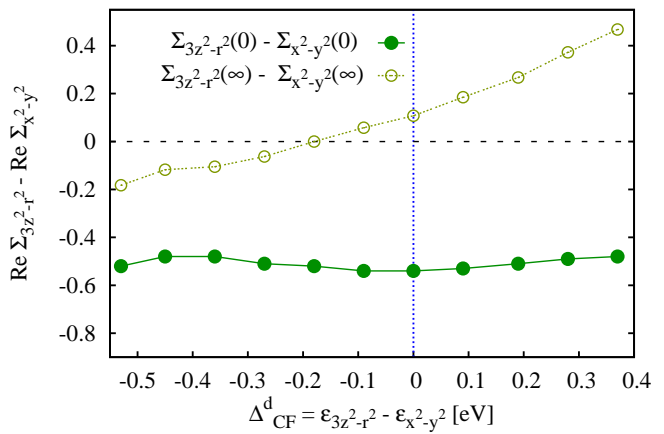} 
        \caption{\label{Fig5_dp}  Difference of the real part of the DMFT
	    self-energies  for the two $d$-orbitals of the four-band  $dp$
	    model with
          $U'= 8$eV, $J=1$eV ($U=10$eV),  $\beta=100$eV$^{-1}$ and the filling is $n_{tot}=5$. Shown is the low and high frequency asymptotes, i.e. the   extrapolation to $\omega_n \rightarrow
          0$ (solid symbols) and $\omega_n \rightarrow +\infty$
          (empty symbol,  Hartree contribution to the self-energy).}
\end{figure}

\begin{figure*}[t!]
         \centering\includegraphics[width=0.9\textwidth,angle=0]{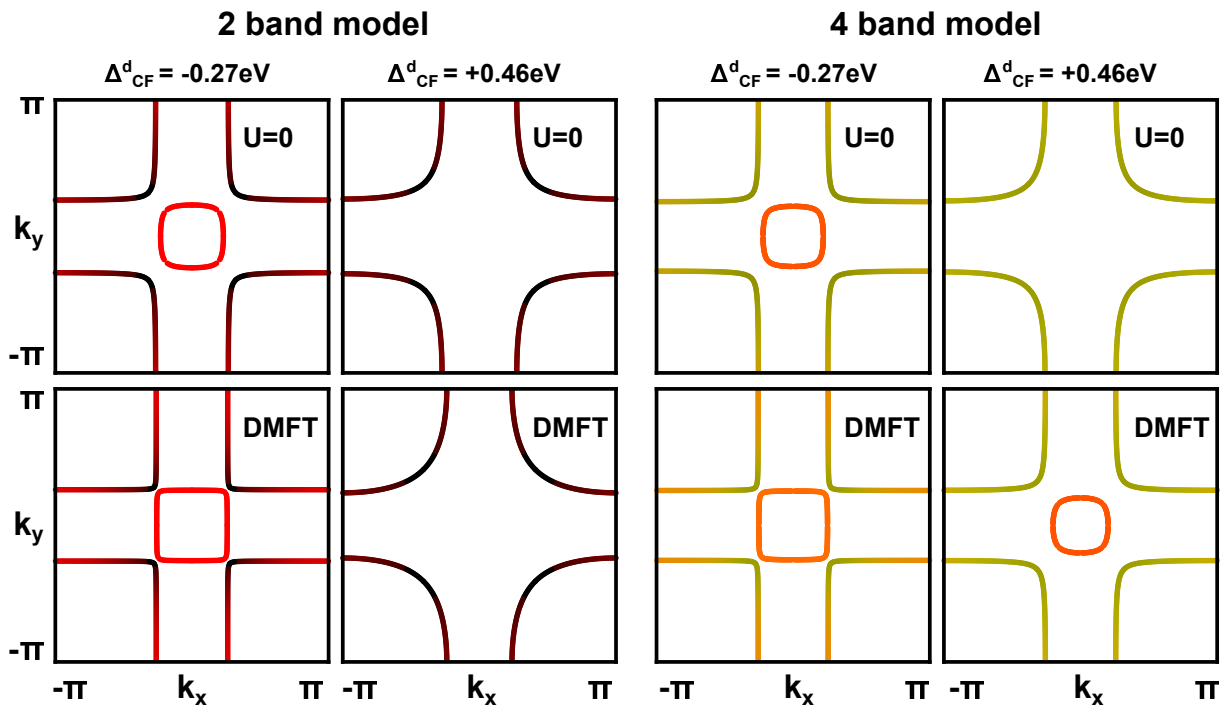} 
        \caption{\label{Fig6_dp} Shape and orbital character of the electronic Fermi surfaces in the two-different basis-sets (left: $d$-only model; right: $dp$ model). The DMFT results are shown in the upper and the noninteracting case in the lower row for comparison.
The two negative/positive values of $\Delta_{CF}^d$ considered, have been chosen as different as possible, with the requirement that a metallic Fermi surface is still found in the interacting case. The color coding of the orbital character (black: $x^2-y^2$-, red: $3z^2-r^2$-, yellow: $p$-orbital(s), is the same as in Fig.\ \ref{fig01}).}
\end{figure*}

For the 4 band $dp$ model the total occupation is
$n_{tot}= n_p+n_d =5$. Because of the stronger localization of the $d$-orbitals in the $dp$ case, larger values for
$U$, $U'$ and $J$ have been considered: 
We have doubled the value
of $U'$  w.r.t. the calculation discussed in
the previous section, i.e. $U'= 8$eV \cite{note3}. 
Because of the smaller screening effects on the Hund's exchange constant, we have considered a reduced enhancement of $J$, which has been fixed to  $1$eV. As before, the relation $U=U'-2J$ holds. 

In Fig. \ref{Fig4_dp} we consider first the occupation of the two $d$-orbitals as a function of the initial
crystal field splitting, starting from the
non-interacting case. Let us recall here that, because of the downfolding procedure, the numerical values of the energy splitting between the $e_g$-orbitals in the one-particle Hamiltonian are evidently different in the $dp$ and in the $d$-only basis-set. They have been labelled, respectively, with $\Delta_{CF}$, see Eq.\ \ref{eq:dpmodel}, and $\Delta_{CF}^d$. Note that the latter includes, as \emph{ligand} field, the effects of $dp$ hybridization. 
To allow for a direct comparison to the effective d-only results we have used the relation between $\Delta_{CF}$ and $\Delta_{CF}^d$ and plot the dp-results as a function of $\Delta_{CF}^d$.

By analyzing the orbital occupation shown in Fig. \ref{Fig4_dp}, the qualitative behavior as a function of $\Delta_{CF}^d$ appears similar as in the $d$-only case. However, an important difference should be noted: In spite of the relatively large
separation ($\sim 2$eV) between $d$- and $p$-bands,
the total occupation of the $d$-orbitals is now much
larger than before ($n_d \sim 1.7 \div 1.8$) due to the $dp$ hybridization. 
Quite remarkably,  according to our DMFT results of Fig. \ref{Fig4_dp},
such an enhanced occupation of the ``correlated'' $d$-orbitals essentially survives also  upon switching on the local interaction.  This fact has an obvious impact on the final results for the orbital
polarization $P$, as now the sign of its change w.r.t.\ the
non-interacting case is no longer related to $\Delta_{CF}^d$:
Fig. \ref{Fig4_dp} shows that
independently on the sign $\Delta_{CF}^d$, one {\sl always} observes a
reduction of the value of $P$, i.e. a net enhancement (reduction) of
the occupation of the  $3z^2-r^2$- ($x^2\!-\!y^2$-) orbital
driven by the electronic interaction.

The analysis of the corresponding
self-energies provides a further confirmation: 
Fig.~\ref{Fig5_dp} shows the plot of the zero frequency extrapolation of the
real part of the self-energy (compare to \ref{Fig3_dp}).
In contrast to the $d$-only results now the interaction correction always reduces the initial crystal field,  
in agreement with the systematic depletion of the
$x^2\!-\!y^2$-orbital (reduction of $P$) observed in the whole parameter
range considered. 

On the basis of these results, it is interesting to examine the consequences for the shape of the Fermi surfaces (FS) in the different cases.
In Fig.~\ref{Fig6_dp} we show the Fermi surfaces for two different values of $\Delta_{CF}^d$ for both the $d$-only and the $dp$ basis-sets. In the upper row we show the non-interacting result and in the lower row the DMFT one. In particular, the two values of $\Delta_{CF}^d$ have been chosen to present the extreme cases 
within our data range (provided that the solution is metallic and has a FS).
As for the analysis of Fig. \ref{Fig6_dp}, we start by considering the non-interacting FS of the upper row for the two different basis-sets: as a consequence of the L\"owdin downfolding, the shape of non-interacting FS corresponding to the same values of $\Delta_{CF}^d$ coincide, while the orbital character encoded by the colors differs in that the $p$ contribution is explicitly present in the $dp$ case. 
Additionally, for each case considered, the contribution to the FS of the $3z^2-r^2$-orbital (encoded by the red color and responsible for the formation of the ``cylindric''-shaped FS sheet around the $\Gamma$-point) is stronger for negative values of $\Delta_{CF}^d$ than for positive. 

By looking at the DMFT data (lower row) for the $dp$ 4-band model, we see that the overall negative correction to $\Delta_{eff}^d$ reported in Fig.\ \ref{Fig5_dp} gives a definite trend for the Fermi surfaces: In the $dp$ case, the interaction, {\sl independently} of the sign of $\Delta_{CF}^d$ always enhances the $3z^2-r^2$ contribution. This is reflected both in the orbital composition (the interacting FS contain more red) and in the shape (enlargement of the central FS sheet). At the same time, our DMFT data unveil quite different trends for the interacting FS of the $d$-only model: The $3z^2-r^2$ contribution is enhanced  {\sl only} for $\Delta_{CF}^d < 0$, i.e., only if the $3z^2-r^2$ was from the beginning the lowest lying of the $e_g$-orbitals.  In fact, coherently with the behavior of $\Delta_{eff}^d$ (see Fig.\ \ref{Fig3_dp}), for positive values of $\Delta_{CF}^d$, the interaction further reduces the  $3z^2-r^2$ contribution to the FS, which becomes progressively ``darker'' colored and more ``$x^2-y^2$''-shaped. Let us note that, depending on the details of the dispersion of the specific $d$-only problem, there are cases in which, for $\Delta_{CF}^d >  0$ the non-interacting FS still presents two-sheets, but the reduction of the  $3z^2-r^2$-character predicted by DMFT, determines a transition to a single sheet FS in the interacting case, analogous to the one calculated in Ref. \onlinecite{Hansmann2009}.

Summarizing the main outcome of the comparsion of our DMFT calculations in different basis-sets for orbital occupations, crystal field corrections and Fermi surface evolution, an evident {\sl qualitative} disagreement is observed, at least in the region of positive values of the initial crystal field $\Delta_{CF}^d$.
\begin{figure*}[t!]
         \centering\includegraphics[width=1.0\textwidth,angle=0]{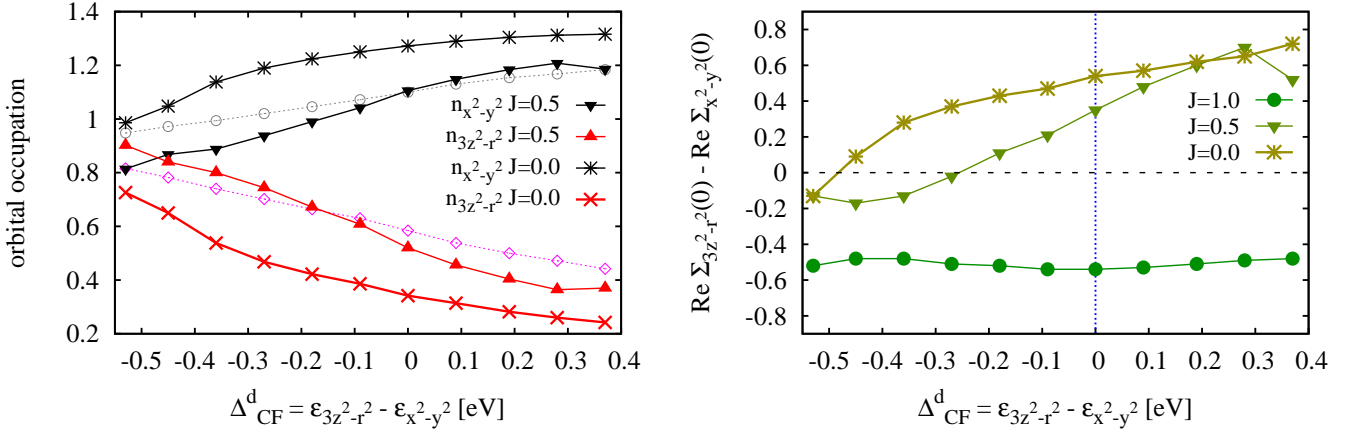}
        \caption{\label{Fig7_dp} Left panel: Orbital occupation of the $d$-orbitals of
          the four-band $dp$  model  as a function of the initial
          crystal field splitting $\Delta_{CF}^d$.
We used  $U'= 8$eV, $\beta=100$eV$^{-1}$, $n_{tot}=5$
and two different values of the Hund's
          exchange $J=0.0, 0.5\,$eV. The DMFT data (solid
          symbols) are compared with the corresponding non-interacting
          results (empty symbols).  Right panel: Corresponding data
          for the difference of the real part of the DMFT self-energies  for
          the two $d$-orbitals of the four-band  $dp$ model extrapolated to $\omega_n \rightarrow 0$
           as a
          function of the initial crystal field splitting
          $\Delta_{CF}^d$. The self-energy data are  also compared with the corresponding DMFT data at
          $J=1.0$, previously shown in Fig. 5.}
\end{figure*}
It is worth noticing that for positive values of $\Delta_{CF}^d$, the trend we found in the $dp$-basis-set is consistent with the results of the $dp$ calculations for the Ni-based  heterostructures of  Ref. \onlinecite{Hanmillis}: There, it was shown, that even when starting from a relatively significant orbital polarization
for the $x^2\!-\!y^2$-orbital at the LDA level, the polarization was always strongly reduced by the interaction, which obviously has bad implications on the
possibility of actually manipulating the Fermi surfaces of the Ni-based
heterostructures in the ``desired'' cuprate-like way \cite{Khal2008,Hansmann2009}.
As a matter of fact,  this discrepancy between $d$-only and $dp$ results has 
 been found here by also including the SU(2) symmetry of the local interaction on the $e_g$-bands in both DMFT calculations. This confirms the
 hypothesis of Ref. \onlinecite{Hanmillis} that such a discrepancy in
 the theoretical predictions does not originate from a different
 treatment of the interactions between the calculations of 
 Refs. \onlinecite{Hansmann2009, Hansmann2010} and \cite{Hanmillis}, but
rather from some more intrinsic difference in the calculations. The most
evident  systematic difference is the
filling $n_d$ of the two $e_g$-orbitals, which,  due to the $dp$ hybridization is strongly increased w.r.t. the quarter-filling level of  the
$d$-only model.  While this is rather obvious at the LDA level, we note
that the  occupation of the $d$ manifold does not change much even 
in presence of the interaction, for typical choices of
the double-counting term for the DMFT (see Sec. II, for details, and
also Ref. \onlinecite{Hanmillis}).  Quite interestingly, the possible role of an 
enhanced $d$-orbital occupation in $dp$ calculations has been also recently
addressed, for the different problem of the occurrence of the
MIT\cite{Lucamillis}.  

The point we make here is that the Hund's exchange $J$ has a strikingly different effect in the $d$-only and in the $dp$ models, due to the fact that, close to half-filling, $J$ drives the system very effectively towards the Hund's rule high-spin ground state \cite{Lucajanus}.
For the orbital polarization this means that $J$ does not play a decisive role for the quarter-filled $d$-only model while it becomes extremely important in the $dp$ model, where $n_d \sim 1.7 \div 1.8$. 
The $d$ and $dp$ results can indeed be reconciled qualitatively if the Hund's coupling $J$ for the $dp$ model is equal to $\approx 0.5$eV or smaller.
This is shown in Fig. \ref{Fig7_dp}, in which the $dp$ calculations have been performed reducing the value of $J$ from
$1.0$eV to $J=0.5$eV, and $J=0.0$eV ($U'=U-2J=8$eV is kept fixed, instead).

The combined analysis of the orbital occupation and of the self-energy
(at $\omega_n \rightarrow 0$)  results shows that the  $dp$ results of
Fig \ref{Fig4_dp} and \ref{Fig5_dp} change qualitatively already for
the case  $J=0.5$eV: the overall trend of strong reduction of occupation of 
the first $x^2\!-\!y^2$-orbital disappears for a large region of values
of the initial crystal field $\Delta_{CF}^d$. In fact, the results already at
$J$=0.5eV would lead to a physical situation
which is qualitatively similar to that
predicted by the $d$-only model. As one can expect, the change with
respect to the previous  $dp$ results becomes even larger when setting
$J=0.0$eV, as both trends of orbital occupation and effective crystal
field become exactly opposite, with the $x^2\!-\!y^2$-orbital occupation
always increased by the interaction irrespectively of the value of the
initial $\Delta_{CF}^d$. The sensitivity of the final $dp$ results on the Hund's coupling is very illustrative and shows the crucial role of the Hund's exchange in this situation\cite{note4}.

\section{Results: The role of the $\mathbf{d}$-orbital occupation}

 From the marked discrepancy between $dp$ and $d$ results,
 discussed in the previous section, a general question naturally
 arises: Under which conditions can one expect 
to obtain a qualitatively similar
DMFT description of the interaction effects for $d$-only and $dp$ calculations?

Let us assume here that we have a very accurate estimate
\cite{note5} of the  local interactions and especially of the Hund's
 exchange $J$. One of the most basic differences between
the $d$-only and the $dp$ calculations examined is the different 
value of the occupation of the $d$-orbitals.  This difference is  observed already at the level of
 the non-interacting model and is, remarkably, not strongly
 affected by the inclusion of the interaction:  For the $d$-only model, the
 DMFT calculations have been performed, as usual, at fixed  filling,
 while the value of $d$-orbital occupation for the $dp$ calculations 
presents moderate oscillations around a much larger value of $n_d \sim
1.75$, quite independently of the parameter set ($U$,$J$,$\Delta_{CF}^d$) considered. 
Since the filling of the {\sl correlated} orbitals is a
crucial factor to drive the system towards a Mott-Hubbard MIT, this will naturally  represent one of the key parameters to be considered when comparing LDA+DMFT calculations on different basis-sets.

\begin{figure*}[t!]
   \centering\includegraphics[width=1.\textwidth,angle=0]{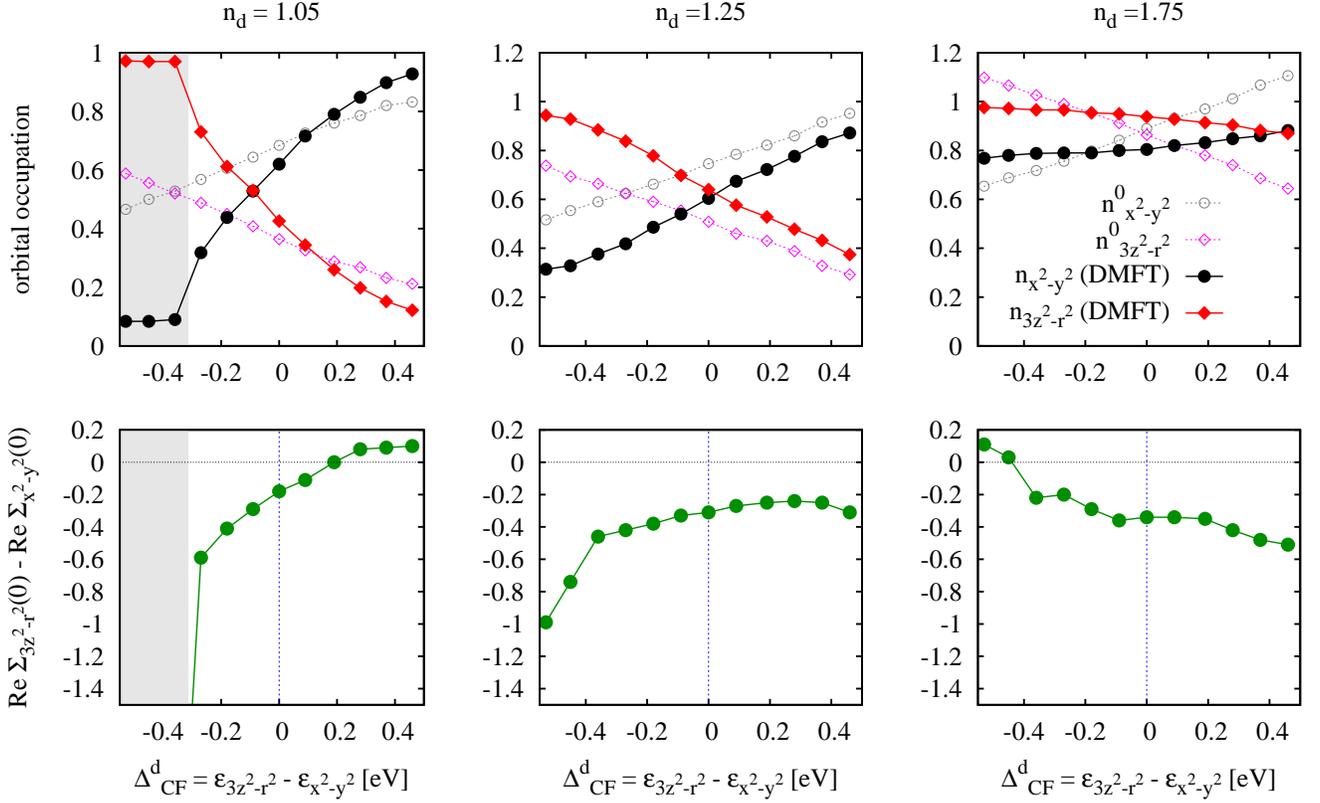}
 \caption{\label{Fig8_dp} Upper row: Orbital occupations of the two orbitals of the
   $d$-only model with $U'= 4$eV, $J=0.75$eV  ($U=5.5$eV),
   $\beta=100$eV$^{-1}$ for different filling $n_{d}=1.05, 1.25, 1.75$ as a function of the initial crystal field splitting
   $\Delta_{CF}^d$. The DMFT data (solid symbols) are compared with the
   corresponding non-interacting results (empty symbols).  Note that a
   filling of $n_d=1.75$ roughly corresponds to the filling of the two
   $d$-orbitals in the $dp$ models considered in the previous sections. Lower row:
   Difference of the real part of the DMFT self-energies  for the two
   orbitals  extrapolated at $\omega_n \rightarrow 0$ (solid symbols)
   as a function of the initial crystal field splitting $\Delta_{CF}^d$
   for the corresponding set of data.}
\end{figure*}

To make our statement more quantitative, we have performed
several additional DMFT calculations for the same $d$-only model as in
Sec. II, but now  varying systematically $n_d$ from the
original quarter-filling level up to the much higher
$n_d \sim 1.75$ found in the $dp$ calculations.
 The results for the orbital occupations and the self-energy at the
 Fermi level are shown in Fig.\ \ref{Fig8_dp} in the
 upper and lower-row panels, respectively. By gradually increasing the total
 occupation of the $d$-orbitals, and focusing on the most interesting
 regime of $\Delta_{CF}^d > 0$, one observes that the parameter region
where the crystal field ``drives'' the final result (i.e. where its magnitude is enhanced by the interaction), shrinks, being confined to
higher and higher values of the initial $\Delta_{CF}^d$.

Looking at the
self-energy plot for $n_d =1.05$,
a  qualitative change in the trend of the effective crystal field
w.r.t. the non-interacting one is found for
$\Delta^d_{*} \sim 0.2$. For $\Delta^d_{CF} < \Delta_{*}^d$ the trend for
the effective crystal field is {\sl opposite} to that observed 
for quarter-filling, i.e. the original crystal field
$\Delta^d_{CF}$  is {\sl reduced} by the interaction. An analogous trend is also observed for the changes in the orbital occupations, w.r.t. the non-interacting ones. 

By further increasing $n_d$ to $1.25$ (second panels of Fig. \ref{Fig8_dp})),  the threshold $\Delta_{*}^d$
is already shifted beyond the border of the parameter region considered
($\Delta_{*}^d > 0.5$). Finally, if one performs the DMFT $d$-only
calculations for $n_d=1.75$, which is roughly similar as in $dp$ model, one indeed finds a similar behavior as for the  $dp$ model:  the interaction correction to the crystal field (Fig. \ref{Fig8_dp}),  overall negative for all positive values of $\Delta_{CF}^d$ considered, resembles  that of the  $dp$ model in  Figs.\ \ref{Fig5_dp}, reflecting the tendency to deplete the $x^2-y^2$-orbital in presence of the electronic interaction. 

For a more general analyis of our data in the whole range of $\Delta_{CF}^d$, let us first recall that, due to the different bandwidths  of the two $e_g$-orbitals, no symmetric behavior
between the regions of positive and negative $\Delta_{CF}^d$  can be expected. 
In particular, in the region $\Delta_{CF}^d < 0$ (i.e., the less relevant one
in the perspective of the Ni-based heterostructures), the crystal field 
``enhancement'' at quarter-filling was much larger  than  the
corresponding one for $\Delta_{CF}^d > 0$ (see Fig. \ref{Fig3_dp}), also
as an effect of the closer proximity of the MIT in this parameter
region.  Hence, while the effects of an increased $n_d$ are always at
work,  here they become first visible as a gradual weakening of
the quarter-filling trends (e.g., as a mitigation of the strongly negative correction to the effective crystal field for $\Delta_{CF}^d < 0$), before getting to an inversion of them:
The gray shadow area, marking the onset of the MIT for one of the two
orbitals i.e., the (almost) half-filled one, is shrinking in 
the $n_d=1.05$ plot, and essentially disappears at $n_d=1.25$. Only at 
higher densities one eventually observes a (weak) sign change of the effective crystal field correction at the lowest border of the parameter region considered  ($\Delta_{CF}^d \sim -0.5$). 

These results demonstrate the essential role played by the density
of the $d$ manifold for determining the
final DMFT results on different basis-sets, 
at least for the important aspect of the Fermi level properties.  The starting value of $n_d$ will decide if, {\sl given a correct ab-initio estimate} of the interaction
parameters of the multi-orbital Hubbard Hamiltonian,
the physics of the interacting system will be {\sl driven} by
the (original) crystal field, or rather by the Hund's rule tendency
for equally occupied orbitals.
Therefore, it will be {\it a priori} quite hard to reconcile DMFT calculations
performed on different basis-sets, without considering the
corresponding occupation of the {\sl correlated} manifolds.

From the perspective of actual material calculations,
the strong dependence of the final LDA+DMFT results for the Ni-based
heterostructures can be put in a rather general framework.  In fact, the quarter-filling
physics particularly favors the crystal field-dominated
physics, since the Hund's exchange is weaker in a system with one electron (on average).
Hence, if the $dp$ hybridization can drive the system away from this
regime (as is the $dp$ model considered here with an average filling of
$n_d \sim 1.75$), the Hund's exchange will easily prevail over the
crystal field effects.
The same interpretation can likely explain, why the results of
LDA+DMFT calculations for the Fe-based superconductors
 do not display, in most cases, such crucial dependence on the basis-set
 considered. There, the five orbitals of the Fe $3d$ manifold 
are characterized by a small value of the {\sl overall} orbital energy splitting $\sim 0.2$eV (in
comparison to the typical values of the order of $1$eV observed in  many transition metal
oxides). This corresponds to a situation of {\sl five} partially filled
($n_d \sim 6$\cite{note_occpnict}) correlated orbitals all very close energetically to the Fermi level, i.e. one of the most favorable playground for strong Hund's exchange processes.
Hence, this case is well inside one of the two regimes, and moderate differences in the orbital occupation of the
$d$ manifold in different basis-sets will {\sl not} result in a qualitative
discrepancy between different LDA+DMFT calculations, with the possible exception of the regimes closest to the Mott transitions\cite{Yin_NatMat,Luca_arXiv,gabiPRL2013}.

\section{Conclusions}

We have thoroughly compared two models for transition metal oxides: a
$d$-only model containing two effective $d$-orbitals and a $dp$ model
 made of two $d$- and two $p$-orbitals. On the single-particle level without
 Coulomb interaction, the two models show the same low-energy physics and band-structure. However if the  Coulomb interaction is taken into account by means of DMFT, this is not the case any more.

The main reason for this discrepancy is the number of 
 $d$-electrons. In the $d$-only  model there  is on average
one $d$-electron per site. For the $dp$ model on the other hand,
the bands are filled with altogether 
$5$ electrons per site: Without hybridization the two
$p$-orbitals at lower energy would be completely filled, and
the $d$-orbitals in the vicinity of the Fermi level would have the
same filling (one electron per site) as for the $d$-only model.
The $d$-orbitals hybridize however with the $p$-orbitals, and, hence,
there is some admixture between the orbitals leading
to a larger filling ($\sim 1.7$) of the $d$-orbitals for the $dp$ model.

This different filling has dramatic consequences for 
the correlated solution of the two models.
For the $d$-only model the Coulomb interaction enhances the
initial crystal field splitting, leading to a situation with 
one $d$-electron in the lowest-lying $d$-orbital.
Depending on the original crystal field (e.g., if $\Delta_{CF}^d > 0$), this can be the $x^2\!-\!y^2$-orbital, which hence dominates the low
energy physics and Fermi surface topology. This kind
of physics can result in a very similar $x^2\!-\!y^2$-shaped Fermi surface in Ni-based heterostructures  as in cuprates.

In contrast, in the $dp$ model the Hund's exchange favors a more even occupation of the two orbitals because the $d$ manifold is much closer to half-filling ($n_d=2$). 
The initial crystal field splitting is therefore reduced and the local moment enhanced \cite{Lucajanus,werner-HS,BHZ-Hubbard}. 
However, if $J$ is low enough this tendency can be weakened recovering qualitatively the results of the $d$-only model (see Fig. \ref{Fig7_dp}).

Vice versa, we have also verified that the $d$-only model displays a similar Hund’s physics as that of the $dp$ model, if we enhance the $d$-electron occupation towards the one of the $dp$ model.
Let us stress that, while an influence of the filling of the correlated manifold on the DMFT results is not surprising in itself, our study identifies its crucial importance for the description of the low-energy physics. In particular, two important aspects have been determined: (i) the rapidity with which qualitative robust trends found in the $d$-only basis-sets can be reversed by slightly varying the density from the ``nominal'' one (e.g, from $n_d=1$ in our case), (ii) the similarity of the interaction effects in different basis-sets observed in presence of a correspondingly similar occupation of the correlated orbitals.

The findings of our model study have important consequences for the
analysis and the interpretation of realistic LDA+DMFT calculations 
for several transition metal oxides and related correlated materials: 
The physical results for $d$-only
LDA+DMFT calculations can indeed dramatically  differ from
 calculations which also include the Oxygen $p$-orbitals.
In fact, since the additional $p$-orbitals lead to a different
$d$-filling, this can determine the aforementioned dramatically different low-energy DMFT physics. Eventually, in the case of contradicting DMFT prediction,  experiments (such as, e.g., ARPES or X-ray
absorption spectroscopy or orbital reflectometry) will show which of the two theoretical set-ups describes the physical reality better.

Furthermore we should note that,  as the
  $dp$ calculation is generally 
far away from an integer filling
of the interacting $d$-orbitals, the occurrence of 
Mott-Hubbard metal-insulator transitions becomes more difficult\cite{Lucamillis,Karstenslides}.
Certainily one can expect non-local correlations beyond DMFT\cite{beyondDMFT}
to be of importance for (quasi) two-dimensional Mott insulators. However,
not showing a Mott insulating phase in other specific cases might well be
a deficit of the $dp$ calculations or, at least,  of the way  
such  $dp$ calculations are performed nowadays, e.g., via the complete neglection of the $dp$ interaction.

Let us emphasize, finally, that in other situations the physics
of the $d$-only and $dp$ calculation can be much more similar. This was e.g. observed in most of the DMFT studies on iron pnictides. In these materials, the presence of a large manifold of partially filled correlated $3d$-orbitals all close to the Fermi level is a common aspect for both  $dp$ and $d$-only calculations. Hence, in both cases the physics is mostly dominated\cite{Pnictides_donly2,Yin_NatMat} by Hund's rule forming a large local magnetic moment in a correlated metallic environment.

\subsubsection*{Acknowledgments}
We acknowledge  financial support from by the Deutsche Forschungsgemeinschaft (DFG) and the Austrian Science Fund (FWF) 
through the Research Units FOR 1346 (FWF I597-N16, A.~T.) and FOR 1162 (DFG, N.~P. and G.~S.) 
and SFB ViCom F41 (K.~H.). We thank O.K. Andersen for exchanging opinions and for his detailed comments about our manuscript, and S.~Biermann M.~Capone, A.~Georges, M.~Haverkort, J.~Kune\v{s},  L.~de~Medici, and P.~Wissgott for fruitful discussions.


\begin{thebibliography}{99}
\bibitem{noteV2O3} We note that such an enhanced crystal field splitting in DMFT has been first observed in realistic calculations for V$_2$O$_3$\cite{V2O3} as well as in model multi-orbital studies\cite{modelmultiband}.
\bibitem{note1}{Note: the completely filled $t_{2g}$-orbitals can affect the
    broadening of the corresponding high-frequency part of the photoemission
    spectra, but {\sl not} the  low-energy electronic structures close to the
    Fermi level \cite{Georgest2g}.}
\bibitem{noteDeltaCFd} Geometrically, it corresponds to a tetragonal distortion, a compression or elongation along one of the cubic C$^4$ axes, of the ligand octahedron.
\bibitem{note2}{Note that the sign of the Hartree
    difference depends also on the value  of the Hund's exchange $J$
    (see also Ref. \onlinecite{GeorgesPROC}). However, for
    plausible values of $J$, the condition $U> 5J$ is typically fulfilled,
    and, hence, the sign prefactor in Eq. \ref{Eq:Hartree} can be usually
    assumed as positive.}
\bibitem{note3}{Note that the trends emerging from our analysis
    have been tested to be quite stable against the changes of $U$, in a
    relatively broad range, while they are more sensible to changes of
    the Hund's exchange $J$,  as discussed in the text.}
\end{thebibliography}
\end{document}